\documentclass[twocolumn]{aastex62}

\usepackage{multirow}

\graphicspath{{}}

\newcommand{\jwst}{{\textsc{JWST}}}
\newcommand{\adl}{{\textsc{AD Leo}}}
\newcommand{\gcr}{{\textsc{GCR}}}
\newcommand{\sep}{{\textsc{SEP}}}
\newcommand{\hep}{{\textsc{HEP}}}
\newcommand{\nox}{{\textsc{NOx}}}
\newcommand{\hox}{{\textsc{HOx}}}
\newcommand{\tise}{TSI$_{\bigoplus}$}
\newcommand{\spe}{{\textsc{SPE89}}} 
\newcommand{\oz}{O$_3$}
\newcommand{\nto}{N$_2$O}
\newcommand{\meth}{CH$_4$}
\newcommand{\chtcl}{CH$_3$Cl}

\newcommand{\pic}[2]{
	\begin{figure}
	 \plotone{#1}
	 \caption{#2\label{#1}}
	\end{figure}}
\newcommand{\picfull}[2]{
	\begin{figure*}
     \plotone{#1}
	 \caption{#2\label{#1}}
	\end{figure*}}

\accepted{June 25, 2018} 
\submitjournal{ApJ}

\shorttitle{Cosmic Ray induced Biosignature Chemistry}
\shortauthors{Scheucher et al.}

\begin{document}

\title{New Insights into Cosmic Ray induced Biosignature Chemistry in Earth-like Atmospheres}


\correspondingauthor{Markus Scheucher}
\email{scheucher@tu-berlin.de, markus.scheucher@dlr.de}

\author{Markus Scheucher}
\affiliation{Zentrum f\"{u}r Astronomie und Astrophysik, Technische Universit\"{a}t Berlin, 10623 Berlin, Germany}

\author{J. L. Grenfell}
\affiliation{Institut f\"{u}r Planetenforschung, Deutsches Zentrum f\"{u}r Luft- und Raumfahrt, 12489 Berlin, Germany}

\author{F. Wunderlich}
\affiliation{Zentrum f\"{u}r Astronomie und Astrophysik, Technische Universit\"{a}t Berlin, 10623 Berlin, Germany}

\author{M. Godolt}
\affiliation{Zentrum f\"{u}r Astronomie und Astrophysik, Technische Universit\"{a}t Berlin, 10623 Berlin, Germany}

\author{F. Schreier}
\affiliation{Institut f\"{u}r Methodik der Fernerkundung, Deutsches Zentrum f\"{u}r Luft- und Raumfahrt,  82234 Oberpfaffenhofen, Germany}

\author{H. Rauer}
\affiliation{Zentrum f\"{u}r Astronomie und Astrophysik, Technische Universit\"{a}t Berlin, 10623 Berlin, Germany}
\affiliation{Institut f\"{u}r Planetenforschung, Deutsches Zentrum f\"{u}r Luft- und Raumfahrt, 12489 Berlin, Germany}


\begin{abstract}
With the recent discoveries of terrestrial planets around active M-dwarfs, destruction processes masking the possible presence of life are receiving increased attention in the exoplanet community. \\
We investigate potential biosignatures of planets having Earth-like (N$_2$-O$_2$) atmospheres orbiting in the habitable zone of the  M-dwarf star \adl. These are bombarded by high energetic particles which can create showers of secondary particles at the surface. 
We apply our cloud-free 1D climate-chemistry model to study the influence of key particle shower parameters and chemical efficiencies of \nox\ and \hox\ production from cosmic rays. We determine the effect of stellar radiation and cosmic rays upon atmospheric composition, temperature, and spectral appearance. 
Despite strong stratospheric \oz\ destruction by cosmic rays, smog \oz\ can significantly build up in the lower atmosphere of our modeled planet around \adl\ related to low stellar UVB. \nto\ abundances decrease with increasing flaring energies but a sink reaction for \nto\ with excited oxygen becomes weaker, stabilizing its abundance. \meth\ is removed mainly by Cl in the upper atmosphere for strong flaring cases and not via hydroxyl as is otherwise usually the case. Cosmic rays weaken the role of \meth\ in heating the middle atmosphere so that H$_2$O absorption becomes more important. We additionally underline the importance of HNO$_3$ as a possible marker for strong stellar particle showers.\\
In a nutshell, uncertainty in \nox\ and \hox\ production from cosmic rays significantly influences biosignature abundances and spectral appearance.
\end{abstract}


\section{Introduction}\label{sec:introduction}

Cool M-dwarf stars are favored targets in exoplanetary sciences due to their high abundance in the Solar neighborhood, a close-in Habitable Zone (HZ), hence short orbital periods, and a high planet/star contrast. For an overview see e.g. \citet{kasting1993,scalo2007, shields2016}. There are however drawbacks. Planets lying in the close-in HZ could be tidally-locked (e.g. \citet{selsis2000,kasting1993}) and could be bombarded by high levels of energetic particles \citep{griessmeier2005}. An additional drawback for M-star planet habitability is the long, bright, pre-main-sequence phase of the parent star, which may devolatilize planets that would later reside in their habitable zones (e.g. \citet{luger2015, ramirez2014, tian2015}). Nevertheless, planets in the HZ of M-dwarf stars could represent the first opportunity to detect atmospheric properties and even biosignatures of rocky extrasolar planets. There are numerous relevant model studies e.g. in 1D \citep{segura2003,segura2005,segura2010,rugheimer2015,kopparapu2013,grenfell2012,tabataba2016} and in 3D \citep{shields2013,shields2016,leconte2013,yang2014,godolt2015,kopparapu2016}.
Interpretation of such potential future observation heavily relies on our detailed understanding of atmospheric physical, chemical, and biological processes and their interaction with different electromagnetic radiation and High Energetic Particles (\hep), such as Galactic Cosmic Rays (\gcr s) and Stellar Energetic Particles (\sep s). The latter has received only limited attention in the exoplanets community so far. Our general understanding of the redistribution of incoming \hep s into secondary particles in so-called air showers through the atmosphere and their influence upon atmospheric chemistry dates back to theoretical work done in the early 1980s \citep{rusch1981, solomon1981}. A more recent study by \citet{airapetian2016} investigated the production of \nto\ via \sep s for the early Earth. While our own star is comparably quiescent, many M-dwarfs show high activities, in which flares with energies comparable to the devastating Carrington event on Earth in the 19th century regularly occur up to a few times a day and orders of magnitude higher energetic events have been observed, e.g. for the here studied M-dwarf \adl\ \citep{atri2017, hawley1991}.\\
Recently, potentially Earth-like planets have been found in the HZ around M-dwarfs (Proxima Cen b, LHS1140 b, and TRAPPIST-1 d-f) which may be studied in further detail with upcoming instrumentation on e.g. the James Webb Space Telescope (\jwst ) \citep{gardner2006} and E-ELT \citep{kasper2010}. The impact of \hep s upon atmospheric chemistry needs further investigation. \hep -induced ion-pairs react with molecular oxygen, molecular nitrogen, and water to cascade into nitrogen oxides (\nox =N+NO+NO$_2$+NO$_3$) and hydrogen oxides (\hox =H+HO+HO$_2$) \citep{porter1976,rusch1981,solomon1981}. \nox\ and \hox\ catalytically destroy ozone (\oz ) in the lower and upper stratosphere respectively \citep{crutzen1970} but can form \oz\ in the troposphere due to the so-called smog mechanism \citep{haagen1952}. They are stored and released from reservoir molecules such as HNO$_3$, depending on e.g. UV radiation and temperature. Recent model studies have shown that \hep\ induced \nox\ and \hox\ from particle showers can indeed significantly reduce \oz\ in an Earth-like atmosphere \citep{grenfell2012, griessmeier2016, tabataba2016}. Production rates of around 1.27 \nox\ \citep{porter1976, rusch1981} and 2.0 \hox\ \citep{solomon1981} per ion pair have been assumed in numerous atmospheric studies, but recent ion-chemistry studies by e.g. \citet{sinnhuber2012} and \citet{verronen2013} have pointed out that the uncertainties in these complex chemical coupling coefficients might be under-estimated, especially when additionally taking into account negative ion chemistry. When conducting numerical studies of rocky planets around active M-dwarfs, such uncertainties can have a major impact on atmospheric abundances - including species influenced by biogenic processes. \oz, for example is removed catalytically by \hox\ and \nox. Also, \meth\ is usually removed by OH, a member of the \hox\ family.\\ 
Based on the above, the main motivation of this work is to compare the influence of different M-dwarf stellar flaring energies to that of the uncertainties in atmospheric \nox-\hox\ production efficiencies from incoming \sep s and show their impact on overall climate and spectral features in transit observations. In Section \ref{sec:methods} we describe the models used for this work and motivate the modeled scenarios, in Section \ref{sec:results} we briefly describe our results, before discussing and comparing them to other relevant works in Section \ref{sec:discussion}. Finally, in Section \ref{sec:conclusion} we draw our conclusions.

\section{Modeling Framework}\label{sec:methods}
\pic{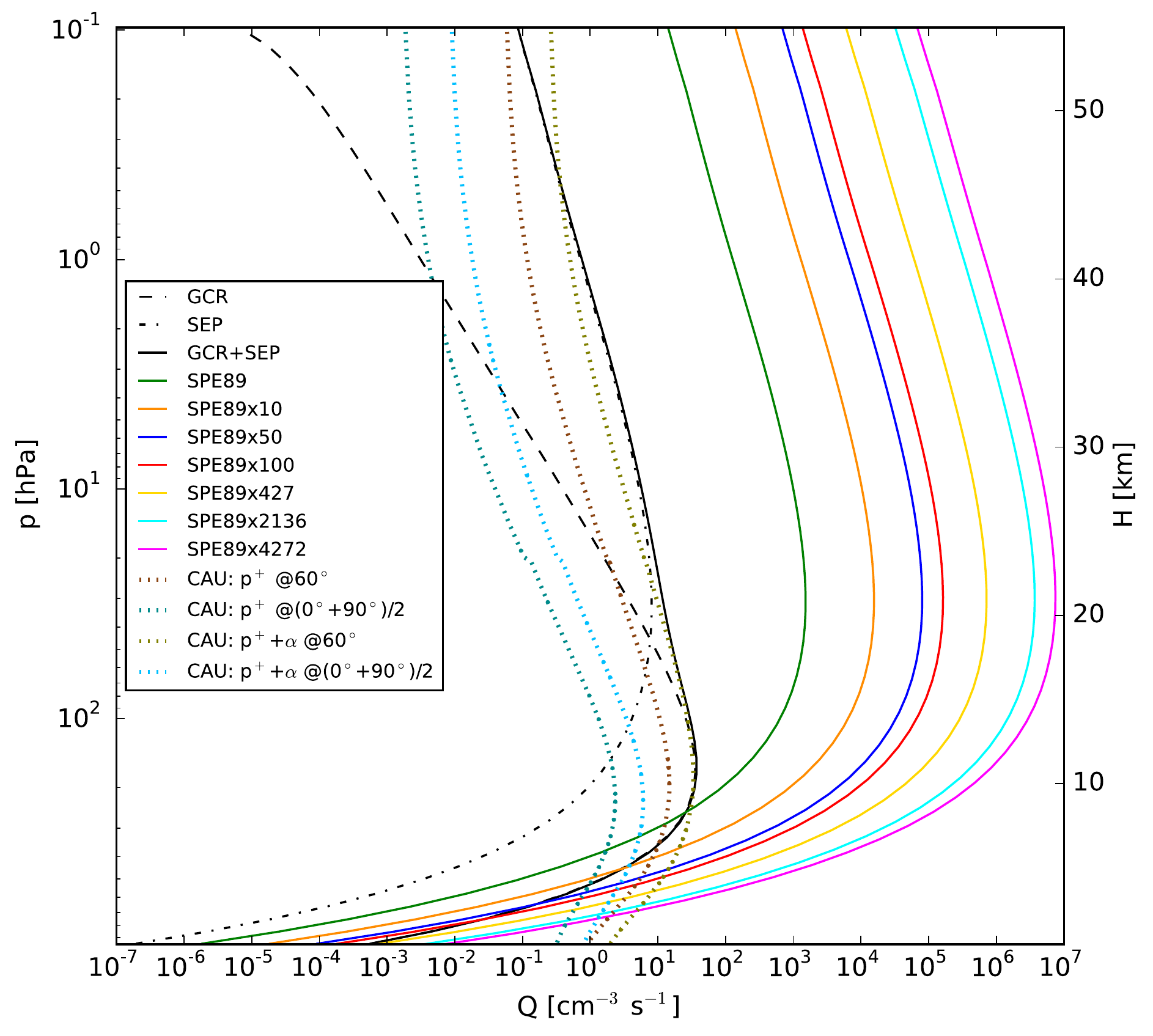}{Ion pair production rate (Q) profiles for different HEP fluxes through Earth's atmosphere calculated in our model (solid). Co-plotted are comparison profiles with \textsc{PLANETOCOSMICS} (CAU) (dotted) for protons (p$^+$) and alpha particles ($\alpha$) and different locations on Earth (60$^\circ$ latitude, or the average of polar (90$^\circ$) and equatorial (0$^\circ$)). Black lines represent solar minimum conditions on Earth from our model, with air showers from \gcr s (dashed), \sep s (dash-dotted), and both combined (solid). The other solid lines represent Q-profiles for a theoretical flaring Sun, shown in multiples of the \spe\ (green), where fluxes in all energy ranges were multiplied by 10 (orange), 50 (blue), and 100 (red). Q-profiles \spe x427 (yellow), \spe x2136 (cyan), and \spe x4272 (magenta), represent separate cases where Earth at 1 AU would receive the same \sep\ flux density as a virtual Earth around \adl\ at a distance of 0.153 AU with stellar flaring strengths of \spe x10 (orange), \spe x50 (blue), and \spe x100 (red), respectively.}
\begin{deluxetable}{|c|c|c|c|c|}
    \tabletypesize{\footnotesize}
    \tablecolumns{5}
    \tablecaption{Model scenarios in this work. 'Star' indicates here the host star of our planet, 'distance' shows the orbital distance of our planet to its host star, and '\hep ' is the energetic particle bombardment on our planet. f$_{\nox}$ and f$_{\hox}$ are the chemical air shower production efficiencies for \nox\ and \hox\ production per ion pair respectively, studied for each configuration.\label{tbl:scenarios}}
    \tablehead{\colhead{Star} & \colhead{Distance [AU]} & \colhead{\hep s} & \colhead{f$_{\nox}$} & \colhead{f$_{\hox}$}}    
    \startdata
        \hline
        \multirow{9}{*}{Sun} & \multirow{9}{*}{1.0} & \gcr & \multirow{9}{*}{\shortstack{1.0\\[4pt]1.27\\[4pt]1.44\\[4pt]1.6\\[4pt]2.0}} & \multirow{9}{*}{\shortstack{0.0\\[4pt]1.0\\[4pt]2.0}}\\
         & & \sep & & \\
         & & \spe & & \\
         & & \spe x10 & & \\
         & & \spe x50 & & \\
         & & \spe x100 & & \\
         & & \spe x427 & & \\
         & & \spe x2136 & & \\
         & & \spe x4272 & & \\
        \hline
        \multirow{6}{*}{\adl} & \multirow{6}{*}{\shortstack{0.153\\[4pt]0.161}} & \gcr & \multirow{6}{*}{\shortstack{1.0\\[4pt]1.27\\[4pt]1.44\\[4pt]1.6\\[4pt]2.0}} & \multirow{6}{*}{\shortstack{0.0\\[4pt]1.0\\[4pt]2.0}}\\
         & & \sep & & \\
         & & \spe & & \\
         & & \spe x10 & & \\
         & & \spe x50 & & \\
         & & \spe x100 & & \\
        \hline 
    \enddata
\end{deluxetable}
\pic{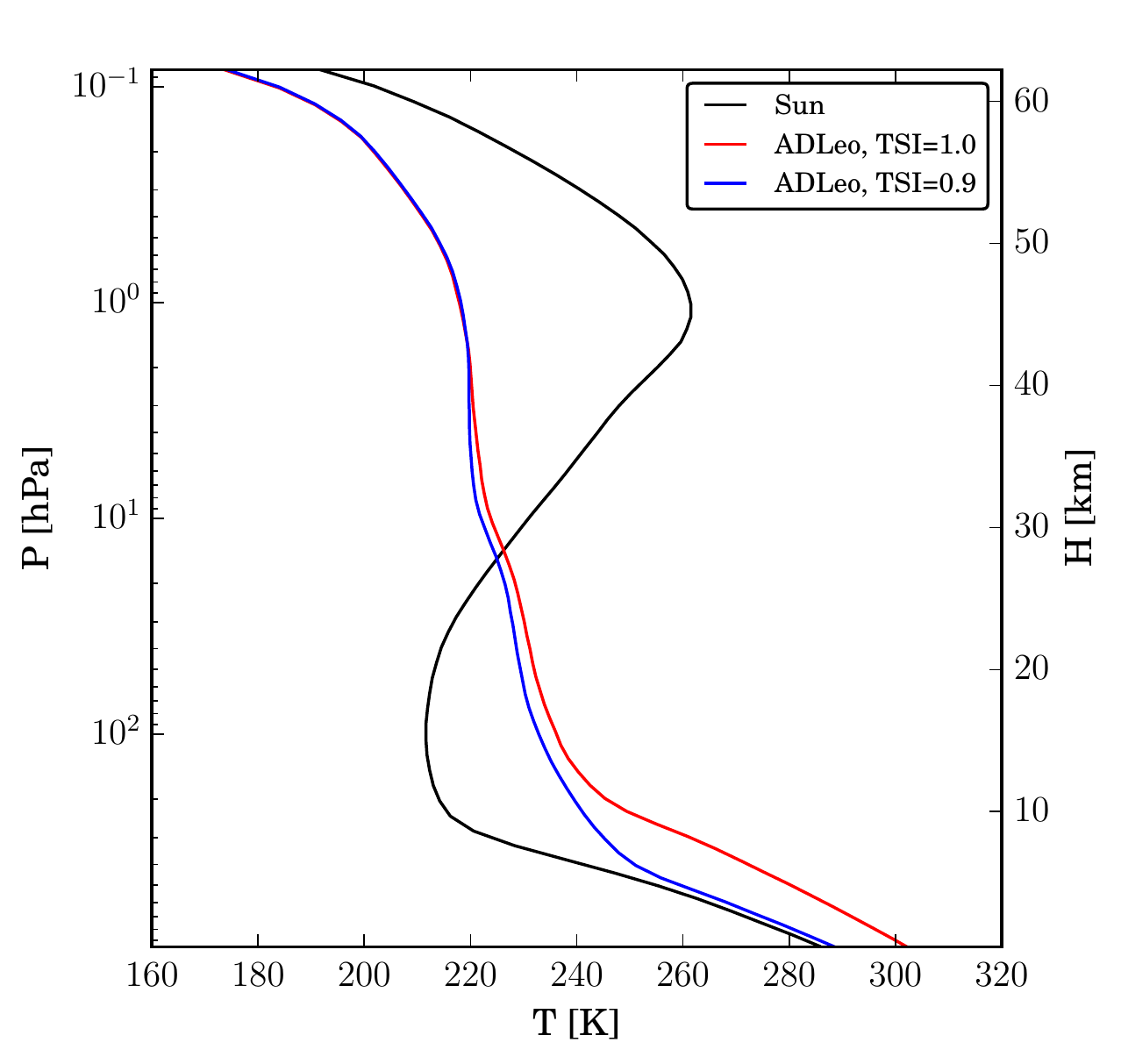}{Temperature profiles for Earth around Sun (black), Earth around \adl\ at a distance of 0.153 AU where TSI = 1.0 \tise\ (red), and Earth around \adl\ at 0.161 AU or 0.9 \tise\ (blue), all with \gcr\ background.}

\subsection{Model description}
We investigate the influence of \gcr s and \sep s from quiescent and flaring stars, on atmospheric chemistry and potential biosignatures like \oz, nitrous oxide (\nto ), \meth, and chloromethane (\chtcl ). We build upon our stationary, global mean, cloud-free coupled climate-chemistry 1D model \citep{tabataba2016}, and update the cosmic rays' propagation.
First we compute the fluxes of primary \hep s, either \gcr s or \sep s, that arrive at the top of our model atmosphere (TOA) at 6.6 Pa, using the magnetospheric model from \citet{griessmeier2005,griessmeier2009}. For Earth reference cases we take \gcr\ and \sep\ measurements outside the Earth's magnetic field for solar minimum conditions and use cut-off energies from \citet{griessmeier2016} for Earth's magnetic shielding. For exoplanet runs we assume the same planetary magnetic field, but scale the HEP fluxes in all energy ranges as follows. For \sep s we use a simplified (conservative) inverse squared scaling with distance from the star (see \citet{grenfell2012}), hence neglect possible magnetic diffusion processes in the heliosphere, which would further increase the power-law exponent beyond 2.0 in the scaling of \hep\ fluxes for short-period orbits. We partly compensate for this effect with higher intrinsic \hep\ flux scenarios from the star (see Model scenarios). For \gcr s at exoplanets we adopt enhanced shielding modulation closer to the star, as described by \citet{griessmeier2009}. Once we have the TOA HEP fluxes, we use the Gaisser-Hillas approach (see e.g. \citet{tabataba2016}) to calculate the atmospheric ion pair production profiles (Q). In the case of Earth, we can compare our approach to more sophisticated models like e.g. the PLANETOCOSMICS Monte Carlo simulations of the cosmic ray induced secondary-particle showers performed by the Christian-Albrechts University (CAU) in Kiel, Germany \citep{fichtner2013}, and find them to be in qualitatively good agreement, as shown in Fig. \ref{Q-Sun-comp__plot.pdf}. Consequently, theses ion pairs cascade into \nox\ and \hox\ species. The \nox-\hox\ production efficiencies describing how many \nox\ and \hox\ are subsequently produced per cosmic rays induced ion pair are widely used in ion chemistry models with values that were calculated by \citet{rusch1981} and \citet{solomon1981}. These values are based on an Earth atmospheric composition assuming ionization to be directly proportional to ionization cross sections, which themselves are taken to be independent of pressure and temperature. Note that negative ion chemistry was taken into account much later by \citet{verronen2013} which further influences the \nox-\hox\ production efficiencies. In our study we made the cosmic rays air-shower parameters in our climate-chemistry model flexible so we can analyze the influence of the uncertainties of these chemical production efficiencies f$_{\nox}$ and f$_{\hox}$ on potential biosignatures compared to the impact of potential stellar flaring scenarios.\\
Lastly, transit spectra i.e. transmission spectra ($\mathcal{T}(\lambda)$) are calculated using the "Generic Atmospheric Radiation Line-by-line Infrared Code" GARLIC \citep{schreier2014} that has been extensively verified \citep[e.g.][]{schreier2018agk} and validated \citep{schreier2018ace}, a FORTRAN90 version of MIRART/SQUIRRL (see e.g. \citet{clarmann2002,melsheimer2005}) used by \citep[e.g. ][]{rauer2011,griessmeier2016,tabataba2016}. In this study we use GARLIC with HITRAN2012 \citep{rothman2013} along with the 'CKD' continua \citep*{clough1989} for calculations of line absorption and Rayleigh scattering parameterization from \citet{sneep2005,marcq2011,murphy1977}. We use temperature, pressure, water vapor, and concentration profiles of those species (23 in total\footnote{Atmospheric species used for transmission spectra in GARLIC are: OH, HO$_2$, H$_2$O$_2$, H$_2$CO, H$_2$O, H$_2$, \oz, \meth, CO, N$_2$O, NO, NO$_2$, HNO$_3$, ClO, CH$_3$Cl, HOCl, HCl, ClONO$_2$, H$_2$S, SO$_2$, O$_2$, CO$_2$, N$_2$}) which are present in both HITRAN2012 as well as in our climate-chemistry model. The corresponding transit depths $\delta(\lambda)$ are calculated using:
\begin{equation}
    \delta(\lambda)=\left(\frac{r_p + h(\lambda)}{r_s}\right)^2,
\end{equation}
where $r_p$ is the planetary radius, $r_s$ the stellar radius, and  $h(\lambda)=\sum_i(1-\mathcal{T}_i(\lambda))\Delta h_i$ is the effective height of the atmosphere for a given wavelength. \\

\subsection{Model scenarios}
In this study we lay focus on virtual Earth-like planets around the active red M4.5 dwarf star AD Leonis, hereafter \adl. In doing so, we place a virtual Earth (1g planet, 1 atm. surface pressure, albedo of 0.21 - tuned to obtain global mean surface temperatures of 288.15 K on Earth) in the HZ around \adl\ in two different positions, starting with the Earth US standard 1976 reference atmosphere (COESA) and allow the climate and chemistry to relax into a new steady state solution. First we start with an Earth around the Sun reference case, using the incoming stellar electromagnetic flux based on \citet{gueymard2004}. To be comparable with earlier studies we next place the planet at a distance of 0.153 AU around \adl, where the Total Stellar Irradiance (TSI) equals the amount the Earth receives from the Sun, before moving it further outwards to 0.161 AU (0.9 \tise) around \adl\ where it receives only 90\% of Earth's TSI. In all \adl\ cases we use the electromagnetic spectrum based on \citet{segura2005}. The former approach of 1.0 \tise\ together with an Earth-like relative humidity profile \citep{manabe1967} leads to surface temperatures larger than 288K (see also scaling arguments in \citet{segura2003} and references therein). Various modeling studies, including early 1D studies \citep[e.g.][]{cess1976, kasting1986}, as well as more recent 3D studies  \citep[e.g.][]{leconte2013,popp2015,godolt2016,fujii2017} have argued that for increased surface temperatures the relative humidity profile may differ from that of the Earth \citep[eg.][]{manabe1967}, which we use here and has also been assumed in previous studies \citep{segura2010,rauer2011,grenfell2012,rugheimer2015b}. Other studies, e.g. \citet{kasting1993} and \citet{kopparapu2013}, assume a fully saturated atmosphere and find that an Earth-like planet around AD Leo would be close to or even inside the inner edge of the habitable zone. This assumption of a fully saturated atmosphere may however overestimate the water concentrations, as shown by 3D studies, see e.g. \citet{leconte2013,yang2014,kopparapu2016}. 3D modeling results by \citet{shields2013} show that an Earth-like planet around \adl\ receiving 90\% insolation may have similar surface temperatures as the Earth around the Sun. Hence, for this case, the assumption of an Earth-like relative humidity profile to determine the water profile in the 1D model seems to be better justified and in line with 3D model results  (see e.g. \citet{godolt2016}). Whereas the approach of placing the planet at 0.161 AU (0.9 \tise) around \adl\ is model dependent, it has the advantage of lying closer to Earth conditions, where our model is validated.\\
For each of the above cases we investigate various stellar activity scenarios for the \hep\ shower through the atmosphere, all based on measurements on Earth, and scaled for \adl\ according to the above mentioned functions. We start with \gcr\ and \sep\ stellar-minimum background cases, as described above. Then we compare various stellar flaring scenarios, all based on GOES 6 and 7 measurements of the medium-hard spectrum flare that hit the Earth in 1989 \citep{smartshea2002}, hereafter \spe. We assume quasi-constant flaring conditions i.e. a flaring frequency faster than the relevant chemical response timescales to investigate long term climate and composition effects of such violent environments rather than short-term variations. We justify our assumption since e.g. Earth's mean \oz\ column does not change significantly with the day-night cycle whereas flaring on \adl\ has been extensively measured to be in the order of a few per day, with event energies exceeding those of the largest recorded events on Earth ($\sim$10$^{32}$erg) \citep{atri2017}. Additionally, from the KEPLER survey we have multiple M-star observations with flaring energies of up to 10$^{36}$erg and frequencies of up to 100 times those of G-stars \citep{maehara2012,shibayama2013,candelaresi2014}. With this in mind, we model higher flaring scenarios by multiplying the \spe\ particle fluxes in all energy ranges by 10, 50, and 100. For \adl\ runs this adds to the ($\sim$40x) enhanced flux density due to proximity to the host star. This means that our virtual Earth at a distance of 0.153AU around an \adl\ flaring with 100 times enhanced \spe\ strength, hereafter \spe x100, actually receives 4272 times the \sep\ flux than Earth received during \spe. For comparison we also investigated an Earth in its current position receiving the same \sep\ flux densities and added these three scenarios as separate cases only for Earth around the Sun cases, flaring with \spe x427, \spe x2136, and \spe x4272 to compare to the \adl\ \spe x10, \spe x50, and \spe x100 cases, respectively. \\
Lastly, for all of the above scenarios, we perform our parameter study and vary the \nox-\hox\ production efficiencies per cosmic rays induced ion pair (Q) within their plausible parameter ranges \citep{rusch1981,solomon1981,sinnhuber2012,verronen2013} for every combination of f$_{\nox}= [1.0,1.27,1.44,1.6,2.0]$ and f$_{\hox}= [0.0,1.0,2.0]$. The full set of model scenarios can be seen in Table \ref{tbl:scenarios}.\\
We would like to remark that in the thought experiment of an Earth-like planet around \adl\ we assume that life would still be present in the form of biogenic surface emissions as on Earth even in the highest flaring cases, despite the hostility from its host star. Also we assume Earth's evolutionary history, and a mean global daytime-average, which might not strictly hold in the case of tidal-locking. We do this for the sake of simplicity to study the impact of \hep s alone. 

\section{Results}\label{sec:results}
\pic{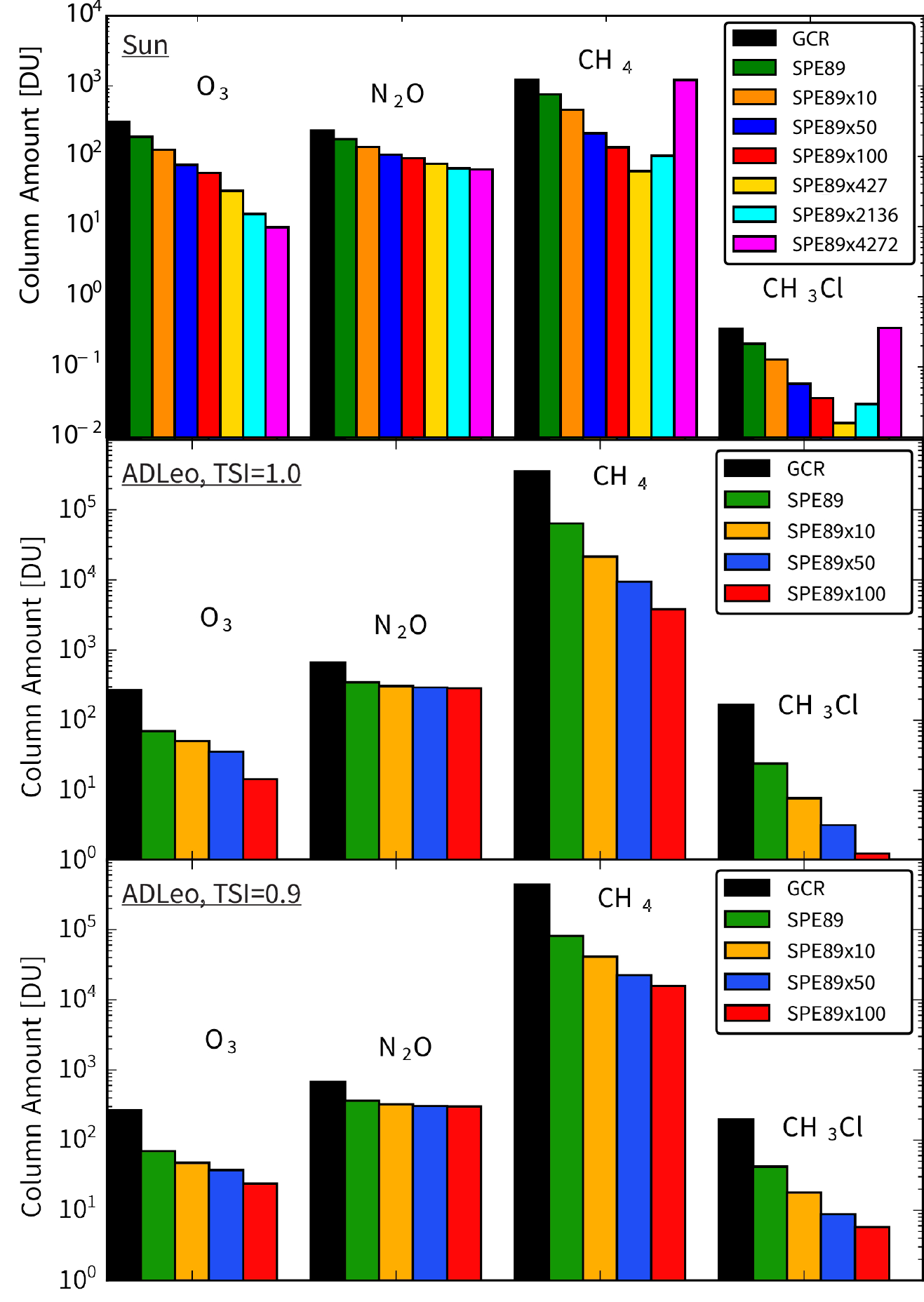}{Atmospheric column amounts for \oz, \nto, \meth, and \chtcl, presented in Dobson units (1 DU = 2.687x10$^{16}$ molec./cm$^{2}$) for Earth around Sun (top), Earth around \adl\ at TSI = 1.0 \tise\ (middle), and Earth around \adl\ at 0.9 \tise\ (bottom). In each panel we show model runs for \gcr\ background (black), stellar flaring scaled from the SPE89 on Earth (green), and the enhanced flaring runs SPE89 multiplied by 10 (orange), 50 (blue), and 100 (red). All particle fluxes received by the planet are scaled from the observed value (@ 1AU) to the appropriate planetary orbital distance. Results for solar minimum \sep\ background are indistinguishable from \gcr\ background, and therefore not shown. The upper panel additionally shows the three cases (\spe x427 (yellow), \spe x2136 (cyan), and \spe x4272 (magenta)) where the Earth around Sun would receive the same \sep\ flux density from the Sun as the virtual Earth around \adl\ at TSI=1.0 \tise\ (middle panel) receives from an \adl\ flaring with \spe x10 (orange), \spe x50 (blue), and \spe x100 (red) strength, respectively (see Sec. \ref{sec:methods} for further explanation).}
\pic{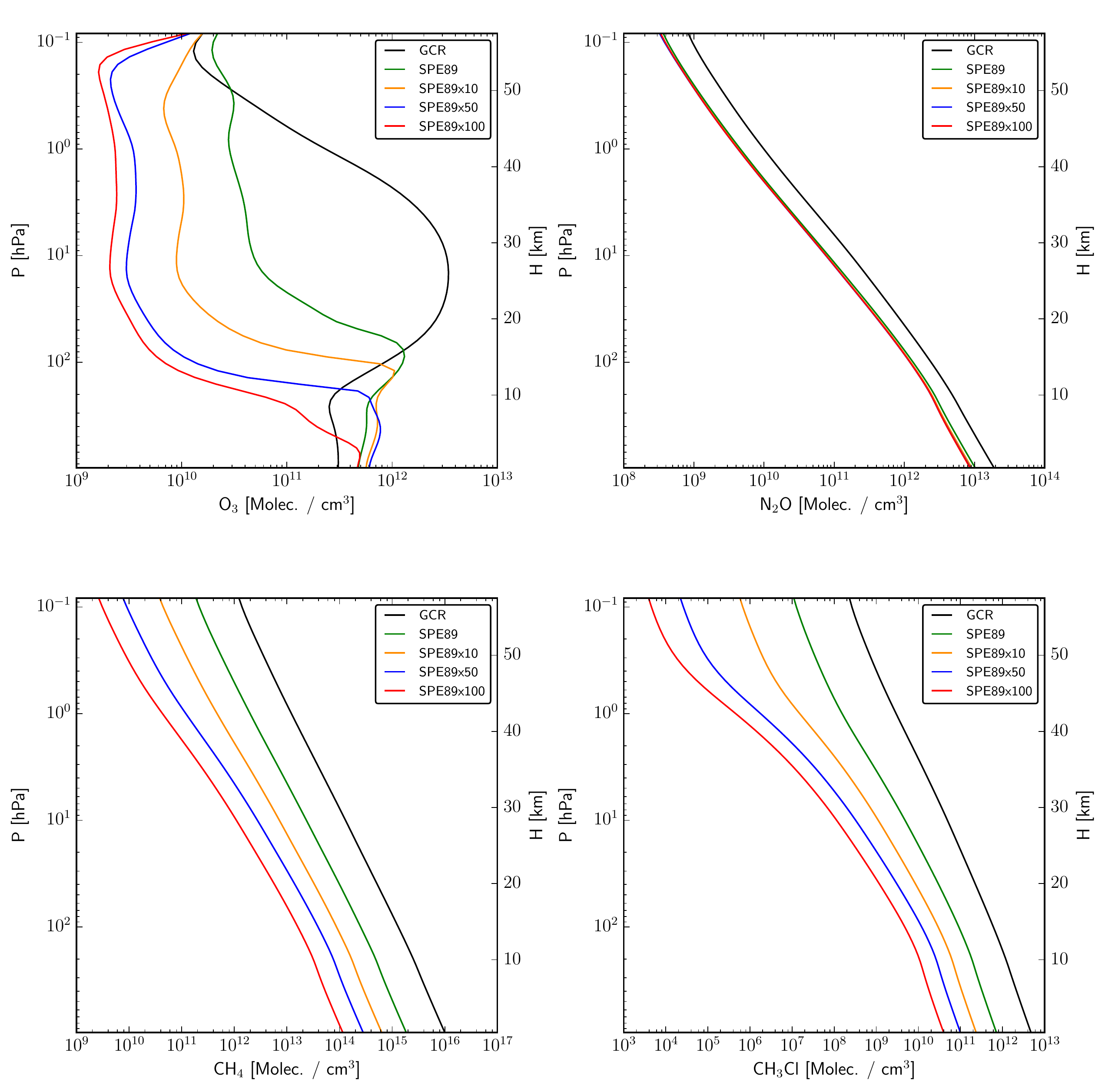}{Molecular profiles for our virtual Earth around \adl\ at 1.0 \tise\ as in Fig. \ref{AllStars.pdf} (middle panel) and red line in Fig. \ref{Temp.pdf}. The ozone profile (upper left), \nto\ (upper right), methane (lower left), and \chtcl\ profile (lower right) are each compared for the different flaring scenarios from Fig. \ref{AllStars.pdf} (middle).}
\picfull{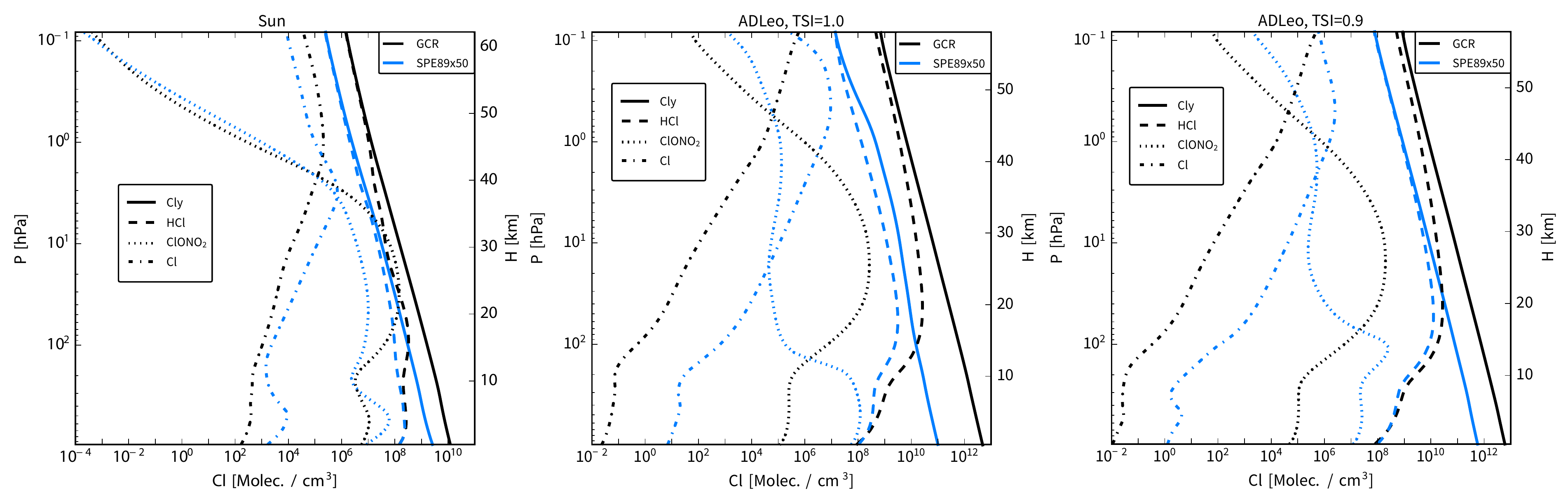}{Atmospheric profiles of several species containing chlorine for the Earth around Sun case (left), \adl\ at 1.0 \tise\ (middle), and \adl\ at 0.9 \tise\ (right). For each case we compare runs with \gcr\ background (black) with runs for 50 times enhanced SPE89 flaring cases (blue). We show the sum of all chlorine containing species (Cly) in our model (solid), HCl (dashed), ClONO$_2$ (dotted), and atomic Cl (dash-dotted).}
\picfull{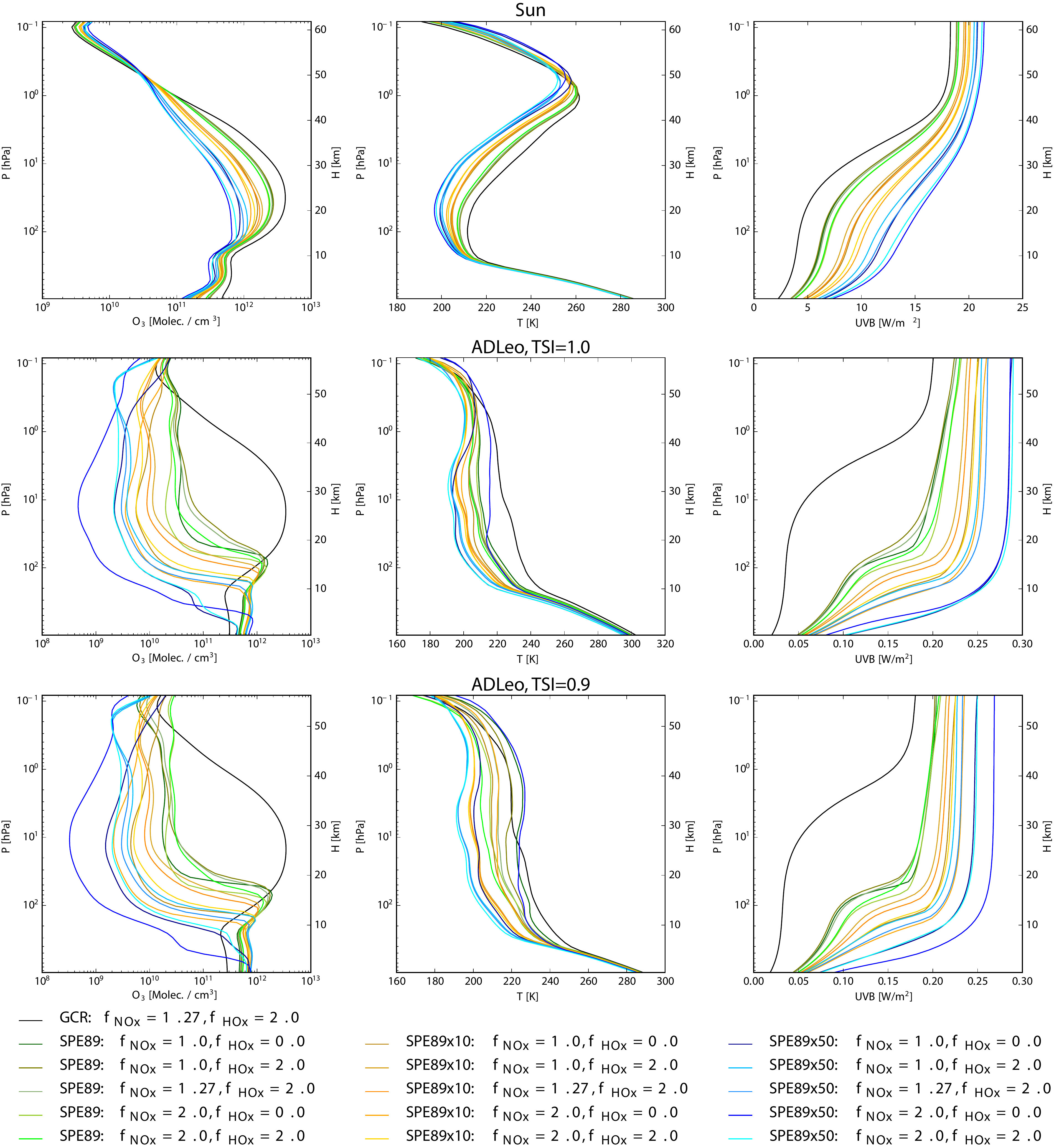}{The influence of chemical \nox-\hox\ fractionation of \hep -induced ion pairs for the flaring scenarios from Fig. \ref{AllStars.pdf}, except the highest flaring cases SPE89x100. The rows show cases for Earth around Sun (top three panels), Earth around \adl\ at 1.0 \tise\ (middle three panels), and Earth around \adl\ at 0.9 \tise\ (bottom three panels). Columns show \oz\ (left), temperature (middle), and UVB (right). The color ranges represent the \gcr\ reference case (black), SPE89 (green colors), SPE89 x 10 (orange colors), and SPE89 x 50 (blue colors). The different shades of green, orange, and blue, represent model runs with different NOx-HOx fractionation for the same respective \hep\ case.}
\picfull{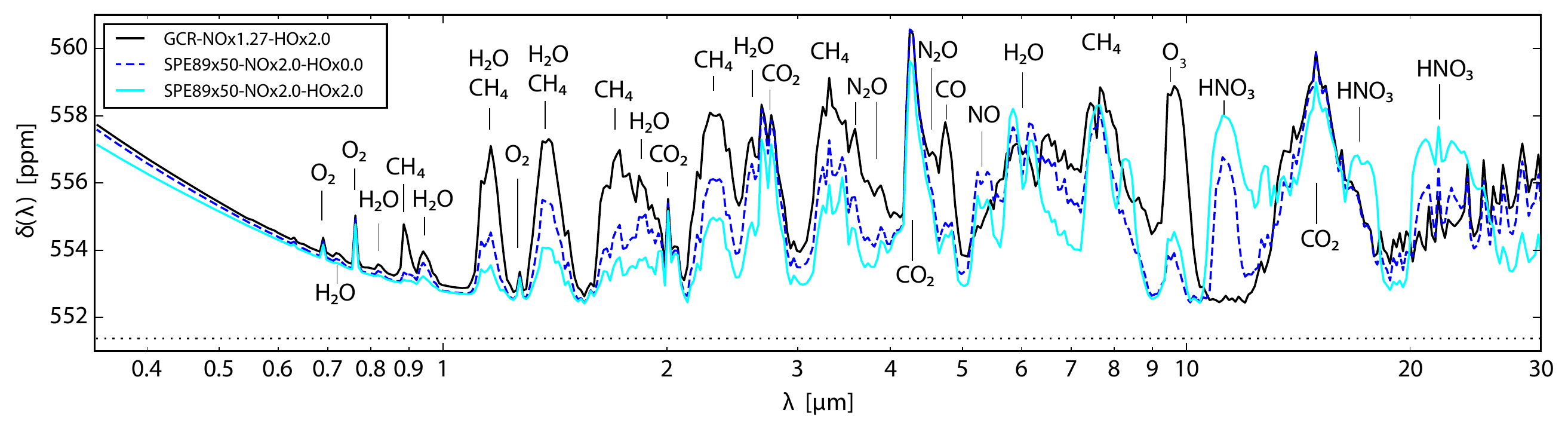}{Modeled transit depths $\delta(\lambda)$ in parts per million over wavelength $\lambda$ (R=100) for a virtual Earth around \adl\ at 0.9 \tise. Shown are three runs from the bottom row of Fig. \ref{AllStarsNoxHox.pdf}: GCR (black), and the last two runs of the high flaring case \spe x50: one with f$_{\nox}=2.0$ and f$_{\hox}=0.0$ (dashed-blue) that shows a similar temperature profile to the \gcr\ case, and one with f$_{\nox}=2.0$ and f$_{\hox}=2.0$ (light-blue) that shows significantly reduced stratospheric temperatures. The dotted black line at the bottom represents the transit depth for the virtual planet without atmosphere (551.3 ppm).}
\pic{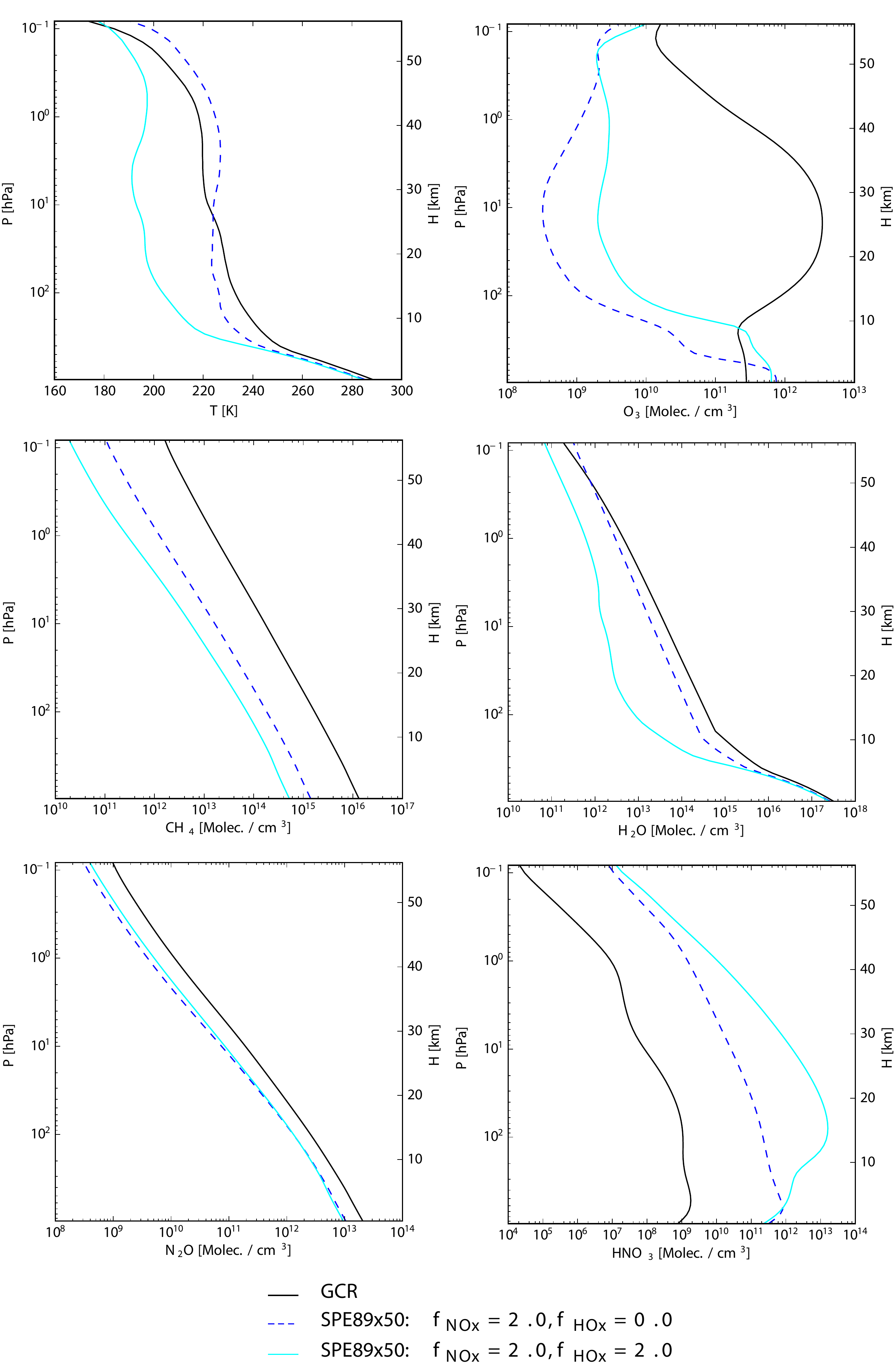}{Atmospheric Profiles of temperature, \oz, \meth, H$_2$O, \nto, and HNO$_3$ for a virtual Earth around \adl\ at 0.9 \tise. Shown are the three runs analyzed in Fig. \ref{trans-spec-wl-con__binned-plot.pdf}: GCR (black), and the last two runs of the high flaring case \spe x50: f$_{\nox}=2.0$ and f$_{\hox}=0.0$ (dashed-blue) and f$_{\nox}=2.0$ and f$_{\hox}=2.0$ (light-blue).}

Figure \ref{Temp.pdf} shows the atmospheric temperature profiles for our three planetary positions, Earth around Sun (black), virtual Earth around \adl\ at a distance of 0.153 AU with a TSI = 1.0 \tise\ (red), and virtual Earth around \adl\ at 0.161 AU with a TSI = 0.9 \tise\ (blue), all with scaled \gcr\ background. The reference is always the \gcr\ background, but solar minimum \sep\ background leads to indistinguishable results for our analysis and is therefore not shown in this work. Until stated otherwise, we show results for the chemical air shower production efficiencies f$_{\nox}$ = 1.27 and f$_{\hox}$ = 2.0, in order to be comparable to other works. For the M-dwarf case at 1.0 \tise\ (red), we calculate an increase in surface temperature to over 300 K mainly due to an enhanced methane greenhouse as found in previous works (e.g. \citet{segura2005}). In addition the stratospheric temperature inversion is essentially non-existent mainly because of the lower near-UV flux compared to the Sun, which reduces \oz\ abundance and heating, as already discussed in previous studies \citep{segura2010,rauer2011}. On moving the planet further away from \adl\ to 0.161 AU (0.9 \tise), the surface temperature in our model reaches Earth-like 289 K, similar to the results of \citet{shields2013}, also lacking the stratospheric temperature inversion. The position of the tropopause (i.e. end of convective regime) in these two \adl\ runs changes significantly from 11 km for the 1.0 \tise\ case to 5.5 km for the 0.9 \tise\ case, compared to $\sim$8.5 km for our Earth around Sun case.\\
Figure \ref{AllStars.pdf} shows atmospheric column amounts of \oz, \nto, \meth, and \chtcl, in Dobson units (1 DU = 2.687x10$^{16}$ molec./cm$^{2}$) for the same planet-star configurations as in Fig. \ref{Temp.pdf}. For each configuration we compare runs with different \hep\ conditions and show the \gcr\ background run (black) as reference. We compare different stellar flaring strengths by multiplying the measured fluxes from \spe\ and scaling the incoming particle flux to the planet's position for the \adl\ runs. The different flaring strength cases vary from \spe x1 to \spe x100 for \adl\ runs and from \spe x1 to \spe x4272 for Earth around Sun runs (see Fig. \ref{AllStars.pdf}). As expected, with increasing flaring strength the \oz\ column is depleted, similar to results of e.g. \citet{grenfell2012} and \citet{tabataba2016}. The same holds qualitatively for \nto\ although note that we calculate here a 'saturation' behavior i.e. where increasing flaring energy has no further effect upon the \nto\ column. Both the \meth\ and \chtcl\ columns also follow the same trend of depletion e.g. from $\sim$1.6 ppmv Earth tropospheric \meth\ concentrations down to $\sim$0.2 ppmv of tropospheric \meth\ in the \spe x100 case. An exception presents the interesting increase in \meth\ and \chtcl\ for our most active Sun cases (upper panel), \spe x2136 (cyan), and \spe x4272 (magenta), which were added for comparison purposes with the \adl\ cases and which we will discuss in Section \ref{sec:discussion}. When we compare the \gcr\ background runs for Earth around Sun with \adl , we see an increase of two orders of magnitude in the \meth\ and \chtcl\ column amounts around \adl\ due to slower \oz\ photolysis resulting in lower OH densities, as discussed in \citet{segura2005}.\\
In order to investigate the column behavior in Fig. \ref{AllStars.pdf}, Fig. \ref{ADL10-molcc__plot.pdf} shows the atmospheric profiles in molecules/cm$^{3}$ in the case of our virtual Earth around \adl\ at 1.0 \tise\ as in Fig. \ref{AllStars.pdf} (middle panel). The color coding for different flaring strengths is the same as in Fig. \ref{AllStars.pdf}. We calculate stratospheric ozone loss (upper left) with increasing influx of \sep s, compared to a quiescent \adl\ (black), as well as increased ozone in the troposphere, which is discussed in Section \ref{sec:discussion}. The \nto\ (upper right) on the other hand, responds only weakly to increasing flare strength, decreasing from $\sim$800 ppbv (2x10$^{13}$ Molec./cm$^3$) (\gcr\ case) to $\sim$250 ppbv (9x10$^{12}$ Molec./cm$^3$) (\spe x100) in the lower atmosphere. Methane and \chtcl\ (lower row) both feature lower concentrations throughout the whole atmosphere with increasing flare strength. The greatly enhanced surface concentration of $\sim$400 ppmv (10$^{16}$ Molec./cm$^3$) \meth\ and $\sim$200 ppbv (5x10$^{12}$ Molec./cm$^3$) \chtcl\ in the \adl\ \gcr\ case (compared to $\sim$1.6 ppmv (3.9x10$^{13}$ Molec./cm$^3$) \meth\ and $\sim$0.5 ppbv (1.2x10$^{10}$ Molec./cm$^3$) \chtcl\ Earth tropospheric concentrations) decreases in our simulations down to $\sim$5 ppmv (10$^{14}$ Molec./cm$^3$) of \meth\ and $\sim$2 ppbv (4x10$^{10}$ Molec./cm$^3$) \chtcl\ for the \adl\ \spe x100 case. \\
In order to investigate the methane response, whose main sinks are OH (lower to mid atmosphere) and Cl (upper atmosphere), we analyze chlorine containing species in our model atmospheres in Fig. \ref{AllStarsCly.pdf}. For all our three planetary configurations we compare background \gcr\ runs (black) with the 50 times enhanced \spe\ flaring cases (blue). The solid lines represent total chlorine (Cl$_y$) i.e. the sum of all chlorine-bearing species, which increases by two orders of magnitude in molecules/cm$^{3}$ when we go from the Earth to an Earth-like planet around \adl. For all cases the majority of stratospheric chlorine is in the form of HCl (dashed), while in the upper stratosphere, atomic chlorine, Cl, also reaches significant levels above 40 km. This is the region where \meth\ production becomes controlled by Cl (dash-dotted), instead of OH, which is the major methane destroying reaction below around 40 km, as discussed in Section \ref{sec:discussion}. ClONO$_2$ (dotted) is only the dominant chlorine-bearing species for a small fraction in Earth's mid-stratosphere for quiescent solar cases.\\
Fig. \ref{AllStarsNoxHox.pdf} compares the influence of varying the chemical air shower production efficiencies f$_{\nox}$ and f$_{\hox}$ for the different flaring cases for the Earth (top), the 1.0 \tise\ \adl\ case (middle), and the 0.9 \tise\ \adl\ case (bottom). We show the resulting ozone profiles (left), temperature (middle), and the UVB environment profile in W/m$^2$ (right). Shades of green represent the \spe\ flaring cases for different \sep\ induced \nox-\hox\ production efficiencies; shades of orange represent the same runs for 10 times enhanced flaring and shades of blue represent the 100 times enhanced flaring cases respectively. Within the green, orange, and blue scenarios, darkest colors represent the lowest parameter values f$_{\nox}$=1.0 and f$_{\hox}$=0.0, while lightest colors represent the highest values of f$_{\nox}$=2.0 and f$_{\hox}$=2.0, with all other combinations in between. For all quiescent star (\gcr) cases (black) all modeled combinations of f$_{\nox}$ and f$_{\hox}$ result in virtually indistinguishable profiles, hence only the cases of f$_{\nox}$=1.27 \nox/Q, and f$_{\hox}$=2.0 \hox/Q, as used by other works, are shown. For the Earth around Sun case, results suggest that the influence of changing the \nox-\hox\ production efficiencies is less important than varying the Sun's flaring strength i.e. amount of incoming \sep s. In the \adl\ cases on the other hand, f$_{\nox}$ and f$_{\hox}$ influence at least the atmospheric temperature and ozone profiles significantly, up to a point where for the 0.9 \tise\ case, a high flaring \adl\ (50 times enhanced \spe) can lead to an atmospheric temperature profile similar to a quiescent \adl\ case, if the chemical air shower parameters through the planet's atmosphere are f$_{\nox}$=2.0 and f$_{\hox}$=0.0. In contrast with f$_{\nox}$=2.0 and f$_{\hox}$=2.0, the temperature may be reduced by up to 40 K throughout almost the whole stratosphere.\\
To investigate whether the above mentioned differences in stratospheric temperature and composition due to different \nox-\hox\ production efficiencies in the \spe x50 strong flaring \adl\ cases with a planet at 0.9 \tise\ (bottom row of Fig. \ref{AllStarsNoxHox.pdf}) have any distinguishable effect on atmospheric spectra, Fig. \ref{trans-spec-wl-con__binned-plot.pdf} shows the corresponding transit depths $\delta(\lambda)$. The spectra are shown with constant spectral resolution R$=\lambda / \Delta\lambda = 100$ over the wavelength range from 0.3-30 $\mu$m. This corresponds to e.g. one mode of the near infrared spectrograph NIRSpec onboard the upcoming \jwst\ mission. We show the contribution of the planetary body i.e. without the atmospheric contribution (dashed-black), which we calculate to be 551.3 ppm. Similar to before, our reference case is the \gcr\ background (black) with f$_{\nox}$=1.27 and f$_{\hox}$=2.0, because runs for all above described combinations of f$_{\nox}$ and f$_{\hox}$ for \gcr\ or \sep\ background cosmic rays result in qualitatively identical atmospheric concentration profiles for all our modeled molecules. First, we compare this to the \spe x50 run for f$_{\nox}$=2.0 and f$_{\hox}$=0.0 (dashed-blue) that results in an almost indistinguishable stratospheric temperature profile. We see the destruction of ozone due to flaring in the weakened 9.6 $\mu$m absorption band as well as reduced absorption by water and methane in the near-infrared (1-2 $\mu$m). We also see HNO$_3$ absorption above 10 microns becoming visible. The second comparison in Fig. \ref{trans-spec-wl-con__binned-plot.pdf} is with the \spe x50 run for f$_{\nox}$=2.0 and f$_{\hox}$=2.0 (light blue). Here we see even stronger suppression of the near-infrared water and methane features, and even stronger HNO$_3$ absorption in the far-infrared. There are also new HNO$_3$ features visible around 17 and 21 $\mu$m. The 9.6 $\mu$m \oz\ absorption band is a little less reduced than in the f$_{\hox}=0.0$ case, due to \nox-\hox\ reactions, that limit the \nox\ sink for \oz, similar to \citet{tabataba2016}. All three runs clearly show the slope due to Rayleigh scattering towards the visible and into the ultraviolet. \\
To further explain the differences in the transit spectra of Fig. \ref{trans-spec-wl-con__binned-plot.pdf}, Fig. \ref{TempStratADL09-molcc__plot.pdf} shows the corresponding atmospheric profiles of temperature, \oz, \meth, H$_2$O, \nto, and HNO$_3$ for the three \adl\ runs at 0.9 \tise using the same color scheme (\oz\ and temperature are also shown and explained in Fig. \ref{AllStarsNoxHox.pdf}). Compared to the \gcr\ case (black) with a fairly constant \meth\ concentration of $\sim$500 ppmv (1.3x10$^{16}$ Molec./cm$^3$) throughout our model atmosphere \meth\ is reduced throughout the whole atmosphere in both \spe x50 cases, f$_{\nox}$=2.0 and f$_{\hox}$=0.0 (dashed-blue) resulting in $\sim$60 ppmv (1.4x10$^{15}$ Molec./cm$^3$) \meth, and f$_{\nox}$=2.0 and f$_{\hox}$=2.0 (light blue) leaving $\sim$20 ppmv (5x10$^{14}$ Molec./cm$^3$) \meth. The H$_2$O profiles show similar behavior, although the H$_2$O abundance is clearly less affected by the \hep s in the \spe x50 f$_{\hox}$=0.0 case, with still $\sim$60-100 ppmv ($\sim$10$^{14}$ Molec./cm$^3$) in the mid and upper atmosphere than in the \spe x50 f$_{\hox}$=2.0 case leaving only $\sim$2-20 ppmv ($\sim$2x10$^{12}$ Molec./cm$^3$) in the mid and upper atmosphere, similar to the according temperature profiles. Again, \nto\ shows only weak responses to both, flaring, and f$_{\hox}$ variation. The HNO$_3$ profiles in both \spe x50 cases are greatly enhanced, as seen in the spectra in Fig. \ref{trans-spec-wl-con__binned-plot.pdf}. In the lower and mid stratosphere HNO$_3$ is further increased in the f$_{\hox}$=2.0 case peaking at $\sim$8 ppmv (2x10$^{13}$ Molec./cm$^3$), which is about ten thousand times the modern Earth's atmospheric concentration, compared to the f$_{\hox}$=0.0 case resulting in only $\sim$90 ppbv ($\sim$3-5x10$^{11}$ Molec./cm$^3$) HNO$_3$, which explains the additional spectral HNO$_3$ features in Fig. \ref{trans-spec-wl-con__binned-plot.pdf}.

\section{Discussion}\label{sec:discussion}
\picfull{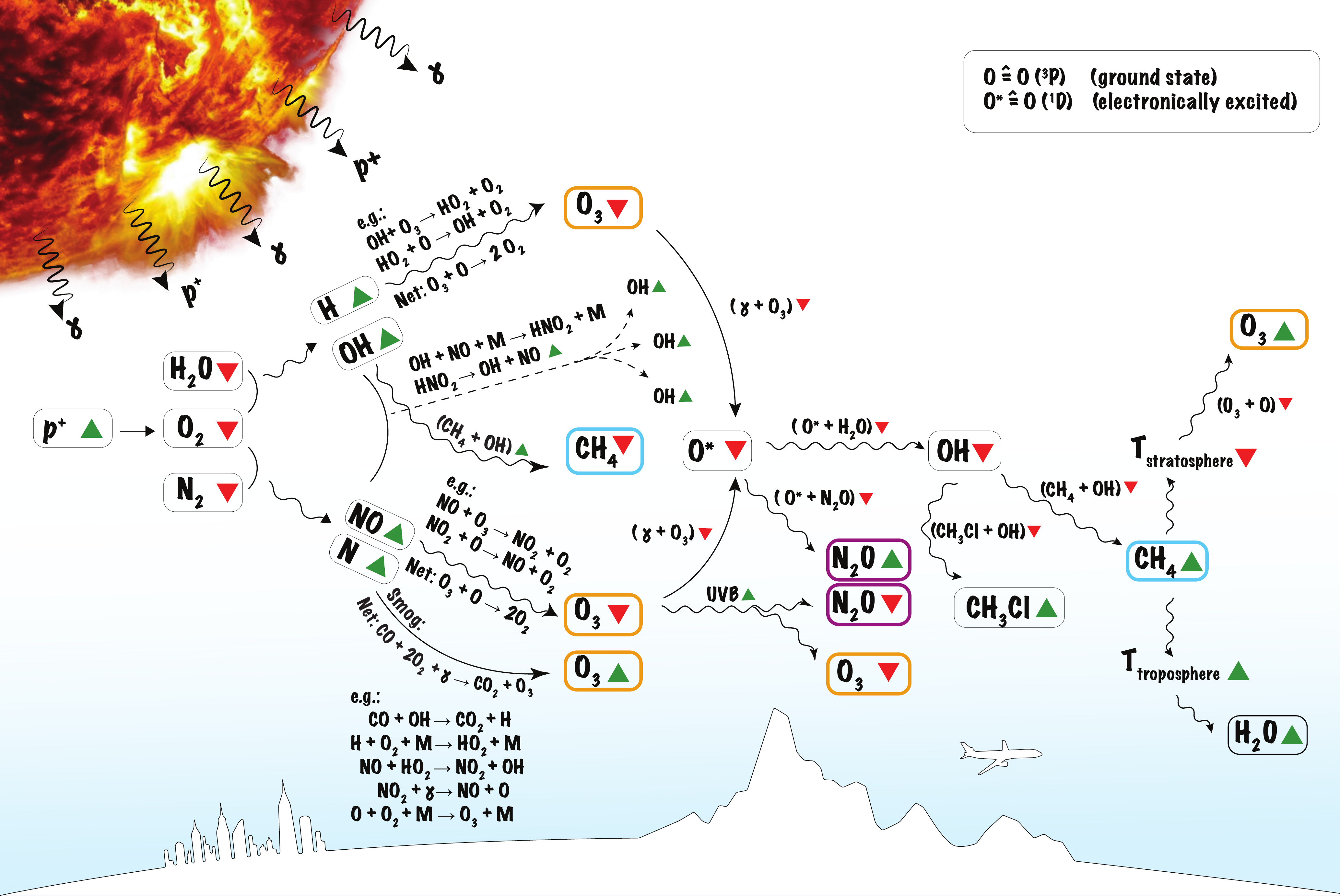}{Schematic diagram showing the biosignature response (simplified) to cosmic rays (p$^+$) and photons ($\gamma$) from the star (upper left corner): Influx of p$^+$ directly decreases (red downward arrows) the abundance of H$_2$O, O$_2$, and N$_2$, by producing (green upward arrows) H, OH, NO, and N. Subsequently the abundance of other molecules also decreases (red arrows) and some increase (green arrows) via several paths and reactions (black). There are several different possible reasons for an e.g. \oz\ or \meth\ increase or decrease, and depending on the host star radiation and p$^+$ activity, some paths are more prominent than others, leading to different abundances. Vertical positions in the diagram provide an indication, where these processes dominate in the atmosphere. Reactions including '+ M' are three-body reactions, while '+ $\gamma$' denotes photolysis. O is the ground state atomic oxygen O($^3$P), and O$^*$ its excited state O($^1$D).}
In this study we analyze how increased \hep\ fluxes associated with active M-stars like \adl, influence the atmospheric chemistry and climate of Earth-like planets in comparison to the impact of chemical production efficiencies. Our special focus lies on biosignatures i.e. species which on Earth are associated with life. Our results in Fig. \ref{Temp.pdf} (red line) show that the surface temperature would increase by over 12 K on placing a virtual Earth-like planet at 1.0 \tise\ around \adl\ when assuming an Earth-like relative humidity profile. However, such a temperature increase would likely change the hydrological cycle leading to a strong water feedback, as shown for a planet around K-stars by \citet{godolt2015}. This could place such planets outside the HZ as suggested by e.g. \citet{kopparapu2013}. Assuming a planet around \adl\ that receives only 90$\%$ \tise\ (blue line), we obtain a moderate surface temperature of 289 K, which is in good agreement with the 3D model results of \citet{shields2013}. \\
Results for different flaring strengths of the same star (Fig. \ref{AllStars.pdf}) suggest stratospheric \oz\ destruction with increasing \hep\ fluxes due to increased catalytic destruction from mainly \nox. All our runs for Sun and \adl\ show this trend. As indicated in Fig. \ref{AllStarsNoxHox.pdf}, the effect of increasing flaring energy upon \oz\ is different for Earth, compared with an Earth-like planet around \adl. While Earth would moderately lose \oz\ in the troposphere and stratosphere, around \adl\ the impact on stratospheric \oz\ is much fiercer, while in the troposphere the enhanced smog \oz\ from \nox\ (see Fig. \ref{Biosignatures-Diagram.pdf} for smog \oz) together with much lower stellar UVB input i.e. less tropospheric \oz\ destruction by photolysis, leads to enhanced tropospheric \oz\ abundances. 
This behavior was confirmed by a detailed analysis of reaction rates of sources and sinks and related pathways. For a theoretically flaring Sun lower stratospheric \oz\ abundances yield higher UVB radiation in the troposphere i.e. higher tropospheric \oz\ photolysis rates, limiting the smog \oz\ build-up. For \adl\ the much lower UVB radiation cannot impact tropospheric smog \oz\ build up efficiently. In the high flaring cases of each host star however, our analysis shows that tropospheric \oz\ undergoes a transition from a smog-dominated build up to a so-called titration limited regime, where the single reaction NO+O$_3$$\rightarrow$NO$_2$+O$_2$ dominates atmospheric \oz\ destruction i.e. efficiently limits \oz\ build up. \\
The main sink for nitrous oxide (\nto) is photolysis, which is generally much slower than \oz\ photolysis. Despite decreased \oz\ for increasing flare energies (Fig. \ref{AllStars.pdf} and \ref{ADL10-molcc__plot.pdf}), hence increased UVB, \nto\ values nevertheless appear to 'saturate'. We can further see this in the UVB profiles in Fig. \ref{AllStarsNoxHox.pdf}: Where stratospheric \oz\ is significantly reduced, UVB absorption in the stratosphere is negligible. The dependence on \oz\ is clearly visible in the \adl\ cases between 18-20 km where \oz\ shows a strong decrease with height. Only the increase in atmospheric density in the troposphere leads to a significant UVB absorption where \oz\ abundances are low. \nto\ is never significantly destroyed by UVB in our study, even where \oz\ is least abundant across all our model runs. \nto\ photolysis is already slower in the stratosphere than \oz\ photolysis,  additionally eddy diffusion might redistribute \nto\ faster than it is photolytically destroyed by UVB and the \nto\ reaction with O($^1$D) becomes O($^1$D) starved hence significantly reduced where \oz\ abundance is lower. Such effects together may explain the \nto\ behavior over a wide variety of flaring conditions. See Fig. \ref{Biosignatures-Diagram.pdf} for an overview.\\
Due to the direct increase of OH (an important \meth\ sink), one would also expect a steady reduction of \meth\ with increasing flare intensity (see Fig. \ref{AllStars.pdf}). For \adl\ cases this holds, but for high flaring Sun cases we see a turning point, above which \meth\ becomes more abundant again. This is a result of decreased OH production. In the troposphere and up to the mid stratosphere, OH is the main sink of \meth. Above that, direct reaction with chlorine becomes the dominant contribution, at least for stellar flaring cases. This effect contributes only weakly to the total column amount however, because of the low number density. In the troposphere, where most of the methane lies, OH is produced by three main sources, the reaction of O($^1$D) with H$_2$O, NO reacting with HO$_2$, and the lesser studied photolysis of HNO$_2$ (see \citet{grenfell1995}) which becomes dominant in the lower stratosphere (this requires future work). O($^1$D) itself is formed almost entirely from \oz\ photolysis. With higher flaring and hence reduced ozone levels, there is a tipping-point, where less O($^1$D) produced from \oz\ leads to a reduction in OH. The same also happens in \adl\ runs, but at higher flaring strengths of around 200 times enhanced \spe\ \sep\ fluxes, due to increased levels of smog \oz\ and lower UVB radiation coming from \adl\ resulting in lower OH production i.e. \meth\ destruction.  Chloromethane, in all cases, follows the methane trends, only with smaller molecular abundances.\\
While the effect of varying f$_{\nox}$ and f$_{\hox}$ is rather weak for the Earth around Sun case, even for an artificially high flaring Sun the effect of varying these parameters for the \adl\ cases becomes very important. The lower row of Fig. \ref{AllStarsNoxHox.pdf} for a virtual Earth around \adl\ at a distance of 0.161 AU i.e. 0.9 \tise\ shows in the high flaring \spe x50 case that especially varying f$_{\hox}$ can result in temperature differences throughout most of the stratosphere of around 40 K. Interestingly enough, the parameter combination of f$_{\nox}=2.0$ and f$_{\hox}=0.0$ around a high flaring \adl\ leads to a temperature profile similar to a quiescent \adl\ while stratospheric ozone abundance in this case is significantly lower (up to four orders of magnitude) than in the quiescent \adl\ cases. \oz\ is known to drive the stratospheric temperature inversion in Earth's atmosphere, but we would like to note here that for low stratospheric \oz\ abundances caused by stellar flaring i.e. reduced or complete lack of temperature inversion, other molecules such as \meth\ and H$_2$O determine stratospheric temperatures as already discussed by e.g. \citet{segura2005,rauer2011,tabataba2016}. We see the main contribution to stratospheric heating in our model from absorption of incoming stellar photons in the near infrared range between 1-2 $\mu$m. As indicated in the transmission spectra in Fig. \ref{trans-spec-wl-con__binned-plot.pdf} H$_2$O and \meth\ bands have overlapping contributions and are therefore hard to distinguish. However, in contrast to the studies mentioned above, in our study of f$_{\nox}$ and f$_{\hox}$ (see Fig. \ref{TempStratADL09-molcc__plot.pdf}) we see a stronger correlation between temperatures and stratospheric H$_2$O profiles and a weakened role of \meth.\\
In the modeled transit spectra in Fig. \ref{trans-spec-wl-con__binned-plot.pdf} we clearly show the effect of reduced ozone, methane, and water for high flaring scenarios when compared to the \gcr\ reference case. All water and methane absorption features are reduced for the flaring case with f$_{\hox}$=0.0, and even more suppressed for the f$_{\hox}$=2.0 case (light blue). Since \nox\ and \hox\ from \hep s can form HNO$_3$ via the reaction NO$_2$+OH+M$\rightarrow$HNO$_3$+M this leads to the absorption features seen in Fig. \ref{trans-spec-wl-con__binned-plot.pdf}. We confirm the HNO$_3$ absorption around 11 microns to be an indicator for an N$_2$-O$_2$ atmosphere exposed to a high flaring stellar environment, as proposed by \citet{tabataba2016}. The second absorption feature of nitric acid around 17 microns is only visible for the case of high \hox\ production from cosmic ray induced ion pairs. Hence, the measurement of this absorption band may be a hint for these values of \hox\ production per ion pair. \citet{rauer2011} and \citet{hedelt2013} already analyzed the telescope time needed with a configuration based on JWST to identify various spectral features. With SNRs derived after \citet{hedelt2013} we estimate that $\sim$35 transits of a theoretical planet around \adl\ (distance to observer 4.9 pc) would be needed to identify the 11 micron HNO$_3$ band with a spectral resolution R=100. The second HNO$_3$ feature around 17 microns might require already a few hundred transits compared to an estimated 3-4 transits with NIRSpec (in the R=100 mode) onboard JWST for the 4.2 micron CO$_2$ band. The third HNO$_3$ absorption feature around 21 microns might be hard to detect at all, as it already lies in the far infrared where the H$_2$O continuum dominates.\\

\section{Conclusions}\label{sec:conclusion}
We have performed atmospheric simulations of virtual Earth-like planets around the flaring M-star \adl\ with our cloud-free 1D climate-chemistry model and have compared the influence of flaring strength with the uncertainty ranges of chemical \nox-\hox\ production efficiencies. \\
New chemical insights found in this work are:
\begin{itemize}
    \item \textbf{\nox-\hox:} The chemical production efficiencies f$_{\nox}$ and f$_{\hox}$ can significantly influence biosignature chemistry and abundances in our model, as well as stratospheric temperatures, and are therefore potentially important for Earth-like planets around M-dwarf stars like \adl. In the Earth's atmosphere, on the other, the influx of \sep s has a much stronger effect than f$_{\nox}$ and f$_{\hox}$, which makes the empirical determination of the latter challenging. 
    
    \item \textbf{HNO$_3$:} Spectroscopic transit measurements of exoplanets may be able to help constrain their stellar environments by looking at e.g. HNO$_3$ features above 10 microns together with infrared \oz, H$_2$O and \meth\ absorption bands. Especially the measurement of the HNO$_{3}$ features at 17 and 21 microns would hint towards high f$_{\hox}$ production.
    
    \item \textbf{Cl:} We introduce and discuss a change of the major \meth\ sink in the stratosphere from OH (lower stratosphere) to Cl (upper stratosphere). This may also become important for worlds with e.g. high volcanic chlorine emissions.
    
    \item \textbf{\oz:} We show that on Earth the UVB radiation from the Sun (G-star) is sufficient to limit global tropospheric smog \oz\ abundances even in hypothetical high flaring Sun scenarios, while we confirm lower atmospheric build-up of \oz\ for Earth-like planets around active M-stars like \adl, as has been modeled in multiple studies e.g. \citet{segura2005,grenfell2012,tabataba2016}.
    
    \item \textbf{\nto:} Atmospheric \nto\ abundance runs into 'saturation' for flaring cases regardless of stellar spectrum, flaring strengths, or stratospheric \oz\ levels. \nto\ reactions e.g. with O($^1$D) in addition to diffusion processes within the atmosphere counteract the \oz\ - UV - \nto\ coupling (See Fig. \ref{Biosignatures-Diagram.pdf}). Hence, destruction of \nto\ by cosmic rays is ineffective in our model.
\end{itemize}
Additionally, in our model OH is the major sink for \meth\ in the lower to mid atmosphere and is directly produced by \sep s, but we find that around high flaring solar-like stars atmospheric \oz\ abundances can significantly drop, which itself is a major source of tropospheric OH production (see Fig. \ref{Biosignatures-Diagram.pdf}). This lack of OH from \oz\ can outweigh OH production from \sep s, subsequently causing unexpectedly high \meth\ abundances (see figures \ref{AllStars.pdf} and \ref{Biosignatures-Diagram.pdf}). Furthermore, in absence of other sources HNO$_2$ can become the main OH source throughout our whole model atmosphere for high flaring host star cases. Further work on this is needed to see for which range of planetary atmospheres HNO$_{2}$ may become important.\\ 
We would like to emphasize once more that \nox\ and \hox\ produced by cosmic rays can become important when studying Earth-like atmospheres around active M-stars.

\acknowledgments
The authors express their thanks to Christina Ciardullo for assistance with graphics.\\
MS acknowledges support from DFG project RA 714/9-1.\\
MG acknowledges support from DFG project GO 2610/1-1.\\
FS acknowledges support from DFG project SCHR 1125/3-1.

\software{GARLIC \citep{schreier2014,schreier2018agk,schreier2018ace}, HITRAN2012 \citep{rothman2013}}
\bibliographystyle{aasjournal}
\bibliography{noxhox}

\begin{thebibliography}{}
\expandafter\ifx\csname natexlab\endcsname\relax\def\natexlab#1{#1}\fi
\providecommand{\url}[1]{\href{#1}{#1}}
\providecommand{\dodoi}[1]{doi:~\href{http://doi.org/#1}{\nolinkurl{#1}}}
\providecommand{\doeprint}[1]{\href{http://ascl.net/#1}{\nolinkurl{http://ascl.net/#1}}}
\providecommand{\doarXiv}[1]{\href{https://arxiv.org/abs/#1}{\nolinkurl{https://arxiv.org/abs/#1}}}

\bibitem[{Airapetian {et~al.}(2016)Airapetian, Glocer, Gronoff, H{\'e}brard, \&
  Danchi}]{airapetian2016}
Airapetian, V., Glocer, A., Gronoff, G., H{\'e}brard, E., \& Danchi, W. 2016,
  Nature Geoscience, 9, 452

\bibitem[{Atri(2017)}]{atri2017}
Atri, D. 2017, Monthly Notices of the Royal Astronomical Society: Letters, 465,
  L34, \dodoi{10.1093/mnrasl/slw199}

\bibitem[{{Candelaresi} {et~al.}(2014){Candelaresi}, {Hillier}, {Maehara},
  {Brandenburg}, \& {Shibata}}]{candelaresi2014}
{Candelaresi}, S., {Hillier}, A., {Maehara}, H., {Brandenburg}, A., \&
  {Shibata}, K. 2014, \apj, 792, 67, \dodoi{10.1088/0004-637X/792/1/67}

\bibitem[{{Cess}(1976)}]{cess1976}
{Cess}, R.~D. 1976, Journal of Atmospheric Sciences, 33, 1831,
  \dodoi{10.1175/1520-0469(1976)033<1831:CCAAOA>2.0.CO;2}

\bibitem[{Clough {et~al.}(1989)Clough, Kneizys, \& Davies}]{clough1989}
Clough, S., Kneizys, F., \& Davies, R. 1989, Atmospheric Research, 23, 229 ,
  \dodoi{https://doi.org/10.1016/0169-8095(89)90020-3}

\bibitem[{{Crutzen}(1970)}]{crutzen1970}
{Crutzen}, P.~J. 1970, Quarterly Journal of the Royal Meteorological Society,
  96, 320, \dodoi{10.1002/qj.49709640815}

\bibitem[{Fichtner {et~al.}(2013)Fichtner, Heber, Herbst, Kopp, \&
  Scherer}]{fichtner2013}
Fichtner, H., Heber, B., Herbst, K., Kopp, A., \& Scherer, K. 2013, in Climate
  and Weather of the Sun-Earth System (CAWSES) (Springer), 55--78

\bibitem[{{Fujii} {et~al.}(2017){Fujii}, {Del Genio}, \&
  {Amundsen}}]{fujii2017}
{Fujii}, Y., {Del Genio}, A.~D., \& {Amundsen}, D.~S. 2017, \apj, 848, 100,
  \dodoi{10.3847/1538-4357/aa8955}

\bibitem[{{Gardner} {et~al.}(2006){Gardner}, {Mather}, {Clampin}, {Doyon},
  {Greenhouse}, {Hammel}, {Hutchings}, {Jakobsen}, {Lilly}, {Long}, {Lunine},
  {McCaughrean}, {Mountain}, {Nella}, {Rieke}, {Rieke}, {Rix}, {Smith},
  {Sonneborn}, {Stiavelli}, {Stockman}, {Windhorst}, \& {Wright}}]{gardner2006}
{Gardner}, J.~P., {Mather}, J.~C., {Clampin}, M., {et~al.} 2006, \ssr, 123,
  485, \dodoi{10.1007/s11214-006-8315-7}

\bibitem[{{Godolt} {et~al.}(2016){Godolt}, {Grenfell}, {Kitzmann}, {Kunze},
  {Langematz}, {Patzer}, {Rauer}, \& {Stracke}}]{godolt2016}
{Godolt}, M., {Grenfell}, J.~L., {Kitzmann}, D., {et~al.} 2016, \aap, 592, A36,
  \dodoi{10.1051/0004-6361/201628413}

\bibitem[{{Godolt} {et~al.}(2015){Godolt}, {Grenfell}, {Hamann-Reinus},
  {Kitzmann}, {Kunze}, {Langematz}, {von Paris}, {Patzer}, {Rauer}, \&
  {Stracke}}]{godolt2015}
{Godolt}, M., {Grenfell}, J.~L., {Hamann-Reinus}, A., {et~al.} 2015, \planss,
  111, 62, \dodoi{10.1016/j.pss.2015.03.010}

\bibitem[{{Grenfell} {et~al.}(1995){Grenfell}, {Bolker}, \&
  {Kleczkowski}}]{grenfell1995}
{Grenfell}, B.~T., {Bolker}, B.~M., \& {Kleczkowski}, A. 1995, Proceedings of
  the Royal Society of London Series B, 259, 97

\bibitem[{{Grenfell} {et~al.}(2012){Grenfell}, {Grie{\ss}meier}, {von Paris},
  {Patzer}, {Lammer}, {Stracke}, {Gebauer}, {Schreier}, \&
  {Rauer}}]{grenfell2012}
{Grenfell}, J.~L., {Grie{\ss}meier}, J.-M., {von Paris}, P., {et~al.} 2012,
  Astrobiology, 12, 1109, \dodoi{10.1089/ast.2011.0682}

\bibitem[{{Grie{\ss}meier} {et~al.}(2009){Grie{\ss}meier}, {Stadelmann},
  {Grenfell}, {Lammer}, \& {Motschmann}}]{griessmeier2009}
{Grie{\ss}meier}, J.-M., {Stadelmann}, A., {Grenfell}, J.~L., {Lammer}, H., \&
  {Motschmann}, U. 2009, \icarus, 199, 526,
  \dodoi{10.1016/j.icarus.2008.09.015}

\bibitem[{{Grie{\ss}meier} {et~al.}(2005){Grie{\ss}meier}, {Stadelmann},
  {Motschmann}, {Belisheva}, {Lammer}, \& {Biernat}}]{griessmeier2005}
{Grie{\ss}meier}, J.-M., {Stadelmann}, A., {Motschmann}, U., {et~al.} 2005,
  Astrobiology, 5, 587, \dodoi{10.1089/ast.2005.5.587}

\bibitem[{{Grie{\ss}meier} {et~al.}(2016){Grie{\ss}meier}, {Tabataba-Vakili},
  {Stadelmann}, {Grenfell}, \& {Atri}}]{griessmeier2016}
{Grie{\ss}meier}, J.-M., {Tabataba-Vakili}, F., {Stadelmann}, A., {Grenfell},
  J.~L., \& {Atri}, D. 2016, \aap, 587, A159,
  \dodoi{10.1051/0004-6361/201425452}

\bibitem[{Gueymard(2004)}]{gueymard2004}
Gueymard, C.~A. 2004, Solar Energy, 76, 423 ,
  \dodoi{https://doi.org/10.1016/j.solener.2003.08.039}

\bibitem[{Haagen-Smit(1952)}]{haagen1952}
Haagen-Smit, A.~J. 1952, Industrial \& Engineering Chemistry, 44, 1342

\bibitem[{{Hawley} \& {Pettersen}(1991)}]{hawley1991}
{Hawley}, S.~L., \& {Pettersen}, B.~R. 1991, \apj, 378, 725,
  \dodoi{10.1086/170474}

\bibitem[{{Hedelt} {et~al.}(2013){Hedelt}, {von Paris}, {Godolt}, {Gebauer},
  {Grenfell}, {Rauer}, {Schreier}, {Selsis}, \& {Trautmann}}]{hedelt2013}
{Hedelt}, P., {von Paris}, P., {Godolt}, M., {et~al.} 2013, \aap, 553,
  \dodoi{10.1051/0004-6361/201117723}

\bibitem[{Kasper {et~al.}(2010)Kasper, Beuzit, Verinaud, Gratton, Kerber,
  Yaitskova, Boccaletti, Thatte, Schmid, Keller, {et~al.}}]{kasper2010}
Kasper, M., Beuzit, J.-L., Verinaud, C., {et~al.} 2010, in Ground-based and
  Airborne Instrumentation for Astronomy III, Vol. 7735, International Society
  for Optics and Photonics, 77352E

\bibitem[{{Kasting} \& {Ackerman}(1986)}]{kasting1986}
{Kasting}, J.~F., \& {Ackerman}, T.~P. 1986, Science, 234, 1383,
  \dodoi{10.1126/science.234.4782.1383}

\bibitem[{{Kasting} {et~al.}(1993){Kasting}, {Whitmire}, \&
  {Reynolds}}]{kasting1993}
{Kasting}, J.~F., {Whitmire}, D.~P., \& {Reynolds}, R.~T. 1993, \icarus, 101,
  108, \dodoi{10.1006/icar.1993.1010}

\bibitem[{{Kopparapu} {et~al.}(2016){Kopparapu}, {Wolf}, {Haqq-Misra}, {Yang},
  {Kasting}, {Meadows}, {Terrien}, \& {Mahadevan}}]{kopparapu2016}
{Kopparapu}, R.~k., {Wolf}, E.~T., {Haqq-Misra}, J., {et~al.} 2016, \apj, 819,
  84, \dodoi{10.3847/0004-637X/819/1/84}

\bibitem[{{Kopparapu} {et~al.}(2013){Kopparapu}, {Ramirez}, {Kasting}, {Eymet},
  {Robinson}, {Mahadevan}, {Terrien}, {Domagal-Goldman}, {Meadows}, \&
  {Deshpande}}]{kopparapu2013}
{Kopparapu}, R.~K., {Ramirez}, R., {Kasting}, J.~F., {et~al.} 2013, \apj, 765,
  131, \dodoi{10.1088/0004-637X/765/2/131}

\bibitem[{{Leconte} {et~al.}(2013){Leconte}, {Forget}, {Charnay}, {Wordsworth},
  \& {Pottier}}]{leconte2013}
{Leconte}, J., {Forget}, F., {Charnay}, B., {Wordsworth}, R., \& {Pottier}, A.
  2013, \nat, 504, 268, \dodoi{10.1038/nature12827}

\bibitem[{Luger \& Barnes(2015)}]{luger2015}
Luger, R., \& Barnes, R. 2015, Astrobiology, 15, 119

\bibitem[{{Maehara} {et~al.}(2012){Maehara}, {Shibayama}, {Notsu}, {Notsu},
  {Nagao}, {Kusaba}, {Honda}, {Nogami}, \& {Shibata}}]{maehara2012}
{Maehara}, H., {Shibayama}, T., {Notsu}, S., {et~al.} 2012, \nat, 485, 478,
  \dodoi{10.1038/nature11063}

\bibitem[{{Manabe} \& {Wetherald}(1967)}]{manabe1967}
{Manabe}, S., \& {Wetherald}, R.~T. 1967, Journal of Atmospheric Sciences, 24,
  241, \dodoi{10.1175/1520-0469(1967)024<0241:TEOTAW>2.0.CO;2}

\bibitem[{Marcq {et~al.}(2011)Marcq, Belyaev, Montmessin, Fedorova, Bertaux,
  Vandaele, \& Neefs}]{marcq2011}
Marcq, E., Belyaev, D., Montmessin, F., {et~al.} 2011, Icarus, 211, 58 ,
  \dodoi{https://doi.org/10.1016/j.icarus.2010.08.021}

\bibitem[{Melsheimer {et~al.}(2005)Melsheimer, Verdes, Buehler, Emde, Eriksson,
  Feist, Ichizawa, John, Kasai, Kopp, {et~al.}}]{melsheimer2005}
Melsheimer, C., Verdes, C., Buehler, S., {et~al.} 2005, Radio Science, 40

\bibitem[{Murphy(1977)}]{murphy1977}
Murphy, W.~F. 1977, The Journal of Chemical Physics, 67, 5877,
  \dodoi{10.1063/1.434794}

\bibitem[{{Popp} {et~al.}(2015){Popp}, {Schmidt}, \& {Marotzke}}]{popp2015}
{Popp}, M., {Schmidt}, H., \& {Marotzke}, J. 2015, Journal of Atmospheric
  Sciences, 72, 452, \dodoi{10.1175/JAS-D-13-047.1}

\bibitem[{{Porter} {et~al.}(1976){Porter}, {Jackman}, \& {Green}}]{porter1976}
{Porter}, H.~S., {Jackman}, C.~H., \& {Green}, A.~E.~S. 1976, \jcp, 65, 154,
  \dodoi{10.1063/1.432812}

\bibitem[{Ramirez \& Kaltenegger(2014)}]{ramirez2014}
Ramirez, R.~M., \& Kaltenegger, L. 2014, The Astrophysical Journal Letters,
  797, L25

\bibitem[{{Rauer} {et~al.}(2011){Rauer}, {Gebauer}, {Paris}, {Cabrera},
  {Godolt}, {Grenfell}, {Belu}, {Selsis}, {Hedelt}, \& {Schreier}}]{rauer2011}
{Rauer}, H., {Gebauer}, S., {Paris}, P.~V., {et~al.} 2011, \aap, 529, A8,
  \dodoi{10.1051/0004-6361/201014368}

\bibitem[{{Rothman} {et~al.}(2013){Rothman}, {Gordon}, {Babikov}, {Barbe},
  {Chris Benner}, {Bernath}, {Birk}, {Bizzocchi}, {Boudon}, {Brown},
  {Campargue}, {Chance}, {Cohen}, {Coudert}, {Devi}, {Drouin}, {Fayt}, {Flaud},
  {Gamache}, {Harrison}, {Hartmann}, {Hill}, {Hodges}, {Jacquemart}, {Jolly},
  {Lamouroux}, {Le Roy}, {Li}, {Long}, {Lyulin}, {Mackie}, {Massie},
  {Mikhailenko}, {M{\"u}ller}, {Naumenko}, {Nikitin}, {Orphal}, {Perevalov},
  {Perrin}, {Polovtseva}, {Richard}, {Smith}, {Starikova}, {Sung}, {Tashkun},
  {Tennyson}, {Toon}, {Tyuterev}, \& {Wagner}}]{rothman2013}
{Rothman}, L.~S., {Gordon}, I.~E., {Babikov}, Y., {et~al.} 2013, JQSRT, 130, 4,
  \dodoi{10.1016/j.jqsrt.2013.07.002}

\bibitem[{{Rugheimer} {et~al.}(2015{\natexlab{a}}){Rugheimer}, {Kaltenegger},
  {Segura}, {Linsky}, \& {Mohanty}}]{rugheimer2015b}
{Rugheimer}, S., {Kaltenegger}, L., {Segura}, A., {Linsky}, J., \& {Mohanty},
  S. 2015{\natexlab{a}}, \apj, 809, 57, \dodoi{10.1088/0004-637X/809/1/57}

\bibitem[{{Rugheimer} {et~al.}(2015{\natexlab{b}}){Rugheimer}, {Segura},
  {Kaltenegger}, \& {Sasselov}}]{rugheimer2015}
{Rugheimer}, S., {Segura}, A., {Kaltenegger}, L., \& {Sasselov}, D.
  2015{\natexlab{b}}, \apj, 806, 137, \dodoi{10.1088/0004-637X/806/1/137}

\bibitem[{Rusch {et~al.}(1981)Rusch, Gérard, Solomon, Crutzen, \&
  Reid}]{rusch1981}
Rusch, D., Gérard, J.-C., Solomon, S., Crutzen, P., \& Reid, G. 1981, \planss,
  29, 767 , \dodoi{http://dx.doi.org/10.1016/0032-0633(81)90048-9}

\bibitem[{{Scalo} {et~al.}(2007){Scalo}, {Kaltenegger}, {Segura}, {Fridlund},
  {Ribas}, {Kulikov}, {Grenfell}, {Rauer}, {Odert}, {Leitzinger}, {Selsis},
  {Khodachenko}, {Eiroa}, {Kasting}, \& {Lammer}}]{scalo2007}
{Scalo}, J., {Kaltenegger}, L., {Segura}, A.~G., {et~al.} 2007, Astrobiology,
  7, 85, \dodoi{10.1089/ast.2006.0125}

\bibitem[{{Schreier} {et~al.}(2014){Schreier}, {Gimeno Garc{\'{\i}}a},
  {Hedelt}, {Hess}, {Mendrok}, {Vasquez}, \& {Xu}}]{schreier2014}
{Schreier}, F., {Gimeno Garc{\'{\i}}a}, S., {Hedelt}, P., {et~al.} 2014, JQSRT,
  137, 29, \dodoi{10.1016/j.jqsrt.2013.11.018}

\bibitem[{Schreier {et~al.}(2018{\natexlab{a}})Schreier, Milz, Buehler, \& von
  Clarmann}]{schreier2018agk}
Schreier, F., Milz, M., Buehler, S., \& von Clarmann, T. 2018{\natexlab{a}},
  J.\ Quant.\ Spectrosc.\ \& Radiat.\ Transfer, 211, 64,
  \dodoi{10.1016/j.jqsrt.2018.02.032}

\bibitem[{Schreier {et~al.}(2018{\natexlab{b}})Schreier, St\"adt, Hedelt, \&
  Godolt}]{schreier2018ace}
Schreier, F., St\"adt, S., Hedelt, P., \& Godolt, M. 2018{\natexlab{b}},
  Molec.\ Astrophysics, 11, 1, \dodoi{10.1016/j.molap.2018.02.001}

\bibitem[{Segura {et~al.}(2005)Segura, Kasting, Meadows, Cohen, Scalo, Crisp,
  Butler, \& Tinetti}]{segura2005}
Segura, A., Kasting, J.~F., Meadows, V., {et~al.} 2005, Astrobiology, 5, 706,
  \dodoi{10.1089/ast.2005.5.706}

\bibitem[{{Segura} {et~al.}(2003){Segura}, {Krelove}, {Kasting}, {Sommerlatt},
  {Meadows}, {Crisp}, {Cohen}, \& {Mlawer}}]{segura2003}
{Segura}, A., {Krelove}, K., {Kasting}, J.~F., {et~al.} 2003, Astrobiology, 3,
  689, \dodoi{10.1089/153110703322736024}

\bibitem[{{Segura} {et~al.}(2010){Segura}, {Walkowicz}, {Meadows}, {Kasting},
  \& {Hawley}}]{segura2010}
{Segura}, A., {Walkowicz}, L.~M., {Meadows}, V., {Kasting}, J., \& {Hawley}, S.
  2010, Astrobiology, 10, 751, \dodoi{10.1089/ast.2009.0376}

\bibitem[{{Selsis}(2000)}]{selsis2000}
{Selsis}, F. 2000, in ESA Special Publication, Vol. 451, Darwin and Astronomy :
  the Infrared Space Interferometer, ed. B.~{Sch{\"u}rmann}, 133

\bibitem[{{Shibayama} {et~al.}(2013){Shibayama}, {Maehara}, {Notsu}, {Notsu},
  {Nagao}, {Honda}, {Ishii}, {Nogami}, \& {Shibata}}]{shibayama2013}
{Shibayama}, T., {Maehara}, H., {Notsu}, S., {et~al.} 2013, \apjs, 209, 5,
  \dodoi{10.1088/0067-0049/209/1/5}

\bibitem[{{Shields} {et~al.}(2016){Shields}, {Ballard}, \&
  {Johnson}}]{shields2016}
{Shields}, A.~L., {Ballard}, S., \& {Johnson}, J.~A. 2016, \physrep, 663, 1,
  \dodoi{10.1016/j.physrep.2016.10.003}

\bibitem[{{Shields} {et~al.}(2013){Shields}, {Meadows}, {Bitz},
  {Pierrehumbert}, {Joshi}, \& {Robinson}}]{shields2013}
{Shields}, A.~L., {Meadows}, V.~S., {Bitz}, C.~M., {et~al.} 2013, Astrobiology,
  13, 715, \dodoi{10.1089/ast.2012.0961}

\bibitem[{{Sinnhuber} {et~al.}(2012){Sinnhuber}, {Nieder}, \&
  {Wieters}}]{sinnhuber2012}
{Sinnhuber}, M., {Nieder}, H., \& {Wieters}, N. 2012, Surveys in Geophysics,
  33, 1281, \dodoi{10.1007/s10712-012-9201-3}

\bibitem[{{Smart} \& {Shea}(2002)}]{smartshea2002}
{Smart}, D.~F., \& {Shea}, M.~A. 2002, Advances in Space Research, 30, 1033,
  \dodoi{10.1016/S0273-1177(02)00497-0}

\bibitem[{Sneep \& Ubachs(2005)}]{sneep2005}
Sneep, M., \& Ubachs, W. 2005, Journal of Quantitative Spectroscopy and
  Radiative Transfer, 92, 293

\bibitem[{Solomon {et~al.}(1981)Solomon, Rusch, Gérard, Reid, \&
  Crutzen}]{solomon1981}
Solomon, S., Rusch, D., Gérard, J., Reid, G., \& Crutzen, P. 1981, \planss,
  29, 885 , \dodoi{http://dx.doi.org/10.1016/0032-0633(81)90078-7}

\bibitem[{{Tabataba-Vakili} {et~al.}(2016){Tabataba-Vakili}, {Grenfell},
  {Grie{\ss}meier}, \& {Rauer}}]{tabataba2016}
{Tabataba-Vakili}, F., {Grenfell}, J.~L., {Grie{\ss}meier}, J.-M., \& {Rauer},
  H. 2016, \aap, 585, A96, \dodoi{10.1051/0004-6361/201425602}

\bibitem[{Tian \& Ida(2015)}]{tian2015}
Tian, F., \& Ida, S. 2015, Nature Geoscience, 8, 177

\bibitem[{Verronen \& Lehmann(2013)}]{verronen2013}
Verronen, P.~T., \& Lehmann, R. 2013, Annales Geophysicae, 31, 909,
  \dodoi{10.5194/angeo-31-909-2013}

\bibitem[{{von Clarmann} {et~al.}(2003){von Clarmann}, {Hopfner}, {Funke},
  {Lopez-Puertas}, {Dudhia}, {Jay}, {Schreier}, {Ridolfi}, {Ceccherini},
  {Kerridge}, {Reburn}, \& {Siddans}}]{clarmann2002}
{von Clarmann}, T., {Hopfner}, M., {Funke}, B., {et~al.} 2003, JQSRT, 78, 381,
  \dodoi{10.1016/S0022-4073(02)00262-5}

\bibitem[{{Yang} {et~al.}(2014){Yang}, {Bou{\'e}}, {Fabrycky}, \&
  {Abbot}}]{yang2014}
{Yang}, J., {Bou{\'e}}, G., {Fabrycky}, D.~C., \& {Abbot}, D.~S. 2014, \apjl,
  787, L2, \dodoi{10.1088/2041-8205/787/1/L2}

\end{thebibliography}

\end{document}